\documentclass[12pt]{article}
\usepackage{graphicx}
\usepackage{amsmath}
\usepackage{amssymb}
\usepackage{caption2}
\begin{document}
\bibliographystyle{prsty}
\begin{center}
{\large {\bf \sc{  Strong decays $B_{s0} \rightarrow B_s \pi$ and
$B_{s1} \rightarrow B^*_s \pi $ with  light-cone QCD sum rules}}} \\[2mm]
Zhi-Gang  Wang \footnote{E-mail,wangzgyiti@yahoo.com.cn.  }    \\
Department of Physics, North China Electric Power University,
Baoding 071003, P. R. China
\end{center}

\begin{abstract}
In this article, we calculate the strong coupling constants
$g_{B_{s0} B_s \eta}$ and $g_{B_{s1} B^*_s \eta}$ with the
light-cone QCD sum rules. Then we take into account the small
$\eta-\pi^0$ transition matrix   according to   Dashen's
 theorem, and obtain the small decay widths for the isospin violation processes $B_{s0}\rightarrow
B_s\eta\rightarrow B_s\pi^0$ and $B_{s1}\rightarrow
B_s^*\eta\rightarrow
 B_s^*\pi^0$. We can search the strange-bottomed
$(0^+,1^+)$ mesons $B_{s0}$ and $B_{s1}$ in the invariant $B_s
\pi^0$ and $B^*_s \pi^0$ mass distributions respectively.
\end{abstract}

PACS numbers:  12.38.Lg; 13.25.Hw; 14.40.Nd

{\bf{Key Words:}}  Strange-bottomed mesons, light-cone QCD sum rules
\section{Introduction}

 Recently, the CDF Collaboration  reports the first observation of two narrow resonances
consistent with the orbitally excited $P$-wave $B_s$ mesons using
$1$ $\mathrm{fb^{-1}}$ of $p\overline{p}$ collisions at $\sqrt{s} =
1.96 \rm{TeV} $ collected with the CDF II detector at the Fermilab
Tevatron \cite{CDF}. The masses of the two states are $M(B^*_{s1})=(
5829.4 \pm 0.7) \rm{MeV}$ and $M(B_{s2}^*) = (5839.7 \pm
0.7)\rm{MeV}$, and they can be assigned as the $J^P=(1^+,2^+)$
states in the heavy quark effective theory \cite{Neubert94}. The D0
Collaboration reports the direct observation of the excited $P$-wave
state $B_{s2}^*$
 in fully reconstructed decays to $B^+K^-$. The mass of the $B_{s2}^*$ meson is measured to be
 $(5839.6 \pm 1.1  \pm 0.7) \rm{MeV}$ \cite{D0}. While the  $B_s$ states with spin-parity
 $J^P=(0^+,1^+)$ are still lack experimental evidence.

The masses of the  $B_s$ mesons with $(0^+,1^+)$ have been estimated
with the potential quark models, heavy quark effective theory and
lattice QCD
\cite{BBmeson1,BBmeson2,BBmeson3,BBmeson4,BBmeson5,BBmeson6,BBmeson7,BBmeson8,BBmeson9,
BBmeson10,Simonov07,Matsuki05,Matsuki07},
the values are different from each other.
 In our previous work \cite{Wang0712}, we study the masses of the  strange-bottomed $(0^+,1^+)$
 mesons with the QCD sum rules,  and observe that the central values
are below the corresponding $BK$ and $B^*K$ thresholds respectively.
The decays $B_{s0}\rightarrow BK$ and $B_{s1}\rightarrow B^*K$ are
kinematically  forbidden.  In previous works,  the  mesons
$f_0(980)$, $a_0(980)$, $D_{s0}$, $D_{s1}$, $B_{s0}$ and $B_{s1}$
are taken as the conventional $q\bar{q}$,  $c\bar{s}$ and $b\bar{s}$
states respectively, and the values of the strong coupling constants
$g_{f_0KK}$, $g_{a_0KK}$, $g_{D_{s0} DK}$, $g_{D_{s1} D^*K}$,
$g_{B_{s0} BK}$ and $g_{B_{s1} B^*K}$ are calculated with the
light-cone QCD sum rules
\cite{WangD1,WangD2,Wang06,Colangelo03,Wang04C,Wang08}. The large
values of the strong coupling constants support the hadronic
dressing mechanism \cite{HDress1,HDress2,HDress3}. Those  mesons may
have small
 $q\bar{q}$,  $c\bar{s}$ and $b\bar{s}$ kernels of the typical
$q\bar{q}$,  $c\bar{s}$ and $b\bar{s}$  mesons size respectively,
strong couplings to the virtual intermediate hadronic states (or the
virtual mesons loops) may result in smaller masses than the
conventional  $q\bar{q}$,  $c\bar{s}$ and $b\bar{s}$ mesons in the
potential quark models, enrich the pure   $q\bar{q}$,  $c\bar{s}$
and $b\bar{s}$ states with other components
\cite{Simonov07,WangD1,WangD2,Wang06,Colangelo03,Wang04C,Wang08,
Swanson06R,Colangelo04R,Klempt07R}.

The $P$-wave heavy mesons  $B_{s0}$ and $B_{s1}$ can decay through
the isospin violation precesses $B_{s0}\rightarrow
B_s\eta\rightarrow B_s\pi^0$ and $B_{s1}\rightarrow
B_s^*\eta\rightarrow
 B_s^*\pi^0$, respectively. The   $\eta-\pi^0$ transition matrix is very small according to
   Dashen's
 theorem \cite{Dashen}$,
 t_{\eta\pi} = \langle \pi^0 |\mathcal {H}
 |\eta\rangle=-0.003\rm{GeV}^2$, they
may be very narrow. In this article, we  calculate the values of the
strong coupling constants $g_{B_{s0} B_s \eta}$ and $g_{B_{s1} B^*_s
\eta}$ with the light-cone QCD sum rules, and study the strong
isospin violation decays $B_{s0} \rightarrow B_s \pi^0$ and $B_{s1}
\rightarrow B^*_s \pi^0 $. In previous work \cite{Zhu06}, the
authors calculate the strong coupling constants $g_{D_{s0}D_s\eta} $
and $g_{D_{s1}D^*_s\eta} $ with the light-cone QCD sum rules, then
take into account the $\eta-\pi^0$ mixing and calculate their pionic
decay widths.

The light-cone QCD sum rules approach carries out the operator
product expansion near the light-cone $x^2\approx 0$ instead of the
short distance $x\approx 0$ while the non-perturbative matrix
elements are parameterized by the light-cone distribution amplitudes
(which classified according to their twists)  instead of
 the vacuum condensates \cite{LCSR1,LCSR2,LCSR3,LCSR4,LCSR5,LCSRreview}. The non-perturbative
 parameters in the light-cone distribution amplitudes are calculated
 with  the conventional QCD  sum rules
 and the  values are universal \cite{SVZ791,SVZ792,Reinders85}.

The article is arranged as: in Section 2, we derive the strong
coupling constants  $g_{B_{s0} B_s \eta}$ and  $g_{B_{s1} B^*_s
\eta}$ with the light-cone QCD sum rules; in Section 3, the
numerical result and discussion; and  Section 4 is reserved for
conclusion.

\section{Strong coupling constants  $g_{B_{s1} B^*_s \eta}$ and $g_{B_{s0} B_s \eta}$ with light-cone QCD sum rules}

In the following, we write down the definitions  for the strong
coupling constants  $g_{B_{s0} B_s \eta}$ and  $g_{B_{s1} B^*_s
\eta}$ respectively,
\begin{eqnarray}
\langle B_{s1}|B^*_s \eta\rangle&=&-ig_{B_{s1} B^*_s \eta}\eta^*
\cdot \epsilon  \, \, , \nonumber\\
\langle B_{s0}| B_s \eta\rangle&=&g_{B_{s0} B_s \eta} \, \, ,
\end{eqnarray}
where the $\epsilon_\mu$ and $\eta_\mu$ are the polarization vectors
of the mesons $B^*_s$ and $B_{s1}$ respectively. The interactions
among the bottomed $(0^-,1^-)$, $(0^+,1^+)$ mesons and the light
pseudoscalar mesons can be described  by the phenomenological
lagrangian \cite{Gatto},
\begin{eqnarray}
\mathcal {L}&=& i h \mbox{Tr} \left[ S_b \gamma_\mu \gamma_5
\mathcal {A}^\mu_{ba} \bar{H}_a\right] +h.c.\, , \nonumber\\
S_a&=&\frac{1+ \!\not\!{v}}{2} \left[B_{a1}^\mu \gamma_\mu \gamma_5
-B_{a0} \right] \, , \nonumber\\
H_a&=& \frac{1+ \!\not\!{v}}{2}\left[B_{a}^{*\mu} \gamma_\mu
-i\gamma_5B_{a} \right] \, , \nonumber\\
\bar{H}_a&=&\gamma^0 H^\dagger_a\gamma^0\, ,\nonumber\\
\mathcal {A}_\mu&=&\frac{1}{2}(L^\dagger\partial_\mu L -R^\dagger\partial\mu R)\, ,\nonumber\\
L&=&R^{\dagger}=\exp[ \frac{i\mathcal {M}}{f_\pi}]\, ,\nonumber\\
\mathcal {M}&=&\left(\begin{array}{ccc}
\frac{\pi^{0}}{\sqrt{2}}+\frac{\eta}{\sqrt{6}}&\pi^{+}&K^{+}\\
\pi^{-}&-\frac{\pi^{0}}{\sqrt{2}}+\frac{\eta}{\sqrt{6}}&
K^{0}\\
K^{-} &\bar{K}^{0}&-\sqrt{\frac{2}{3}}\eta
\end{array}\right) \, ,
\end{eqnarray}
where the $a$ and $b$ are the flavor indexes for the light quarks,
$v^2=1$,  and the $h$ is the strong coupling constant. From the
phenomenological lagrangian, we can obtain $g_{B_{s1} B^*_s \eta}
\propto ih $ and $g_{B_{s0} B_s \eta}\propto h$. The hadronic matrix
elements $\langle B_{s1}|B^*_s \eta\rangle$ and $\langle B_{s0}| B_s
\eta\rangle$  have a relative phase factor $i$, furthermore, we
 take the definition $\langle
B_{s1}|B^*_s \eta\rangle=-ig_{B_{s1} B^*_s \eta}\eta^* \cdot
\epsilon$ as the corresponding one $\langle B_{s1}|B^*
K\rangle=-ig_{B_{s1} B^* K}\eta^* \cdot \epsilon$ in
Ref.\cite{Wang08}, where  a negative sign is chosen   to guarantee
that the strong coupling constant $g_{B_{s1} B^* K}$ has positive
value. The expressions in Eq.(1) are the correct formula, although
there are other definitions \cite{Zhu06}. In literatures, the
super-field $H_a$ are usually defined as $H_a= \frac{1+
\!\not\!{v}}{2} \left[B_{a}^{*\mu} \gamma_\mu -\gamma_5B_{a}
\right]$, the $i$ companied with the pseudoscalar mesons $B_a$ is
missed,  therefor the $i$ in Eq.(1) disappears. Here we take the
correct expression given by A. V. Manohar and M. B. Wise in the book
"Heavy Quark Theory " \cite{Book}.

We study the strong coupling constants  $g_{B_{s1} B^*_s \eta}$
 and $g_{B_{s0} B_s \eta}$ with the
 two-point correlation functions $\Pi_{\mu\nu}(p,q)$ and $\Pi_{\mu}(p,q)$ respectively,
\begin{eqnarray}
\Pi_{\mu \nu}(p,q)&=&i \int d^4x \, e^{-i q \cdot x} \,
\langle 0 |T\left\{J^V_\mu(0) J_{\nu}^{A\dagger}(x)\right\}|\eta(p)\rangle \, , \\
\Pi_{\mu }(p,q)&=&i \int d^4x \, e^{-i q \cdot x} \,
\langle 0 |T\left\{J^5_{\mu}(0) J^{S\dagger}(x)\right\}|\eta(p)\rangle \, , \\
J^V_\mu(x)&=&{\bar s}(x)\gamma_\mu  b(x)\, , \nonumber \\
J^A_{\mu}(x)&=&{\bar s}(x)\gamma_\mu \gamma_5 b(x)\, , \nonumber \\
J^5_{\mu}(x)&=&{\bar s}(x)\gamma_\mu \gamma_5 b(x)\, ,\nonumber  \\
J^S(x)&=&{\bar s}(x) b(x)\, ,
\end{eqnarray}
where the currents $J^V_\mu(x)$, $J^A_\mu(x)$, $J^5_\mu(x)$ and
$J^S(x)$ interpolate the strange-bottomed mesons $B^*_s$, $B_{s1}$,
$B_s$ and $B_{s0}$,  respectively, the external $\eta$ meson has
four momentum $p_\mu$ with $p^2=m_\eta^2$. The $J^5_\mu(x)$ and
$J^A_\mu(x)$ are the same current, we take different notations to
denote that  the contributions from the pseudoscalar meson and
axial-vector meson  are taken  respectively.

 The correlation functions $\Pi_{\mu\nu}(p,q)$ and
$\Pi_{\mu}(p,q)$ can be decomposed as
\begin{eqnarray}
\Pi_{\mu \nu}(p,q)&=&i \Pi_A(p,q) g_{\mu\nu}+i\Pi_{A1}(p,q)p_\mu
q_\nu +i\Pi_{A2}(p,q)p_\nu q_\mu
+ i\Pi_{A3}(p,q)q_\mu q_\nu \, , \nonumber\\
\Pi_{\mu}(p,q)&=&i \Pi_S(p,q) q_{\mu}+i\Pi_{S1}(p,q)p_\mu
\end{eqnarray}
due to the Lorentz invariance. We choose the tensor structures
$g_{\mu\nu}$ and $q_\mu$ for analysis in this article.

According to the basic assumption of current-hadron duality in the
QCD sum rules approach \cite{SVZ791,SVZ792,Reinders85}, we can
insert  a complete series of intermediate states with the same
quantum numbers as the current operators $J^V_\mu(x)$, $J^A_\mu(x)$,
$J^5_\mu(x)$ and $J^S(x)$  into the correlation functions
$\Pi_{\mu\nu}(p,q)$ and $\Pi_{\mu}(p,q)$ to obtain the hadronic
representations. After isolating the ground state contributions from
the pole terms of the mesons $B^*_s$, $B_{s1}$, $B_s$ and $B_{s0}$,
we get the following results,
\begin{eqnarray}
\Pi_{\mu\nu}&=&\frac{\langle0| J^V_{\mu}(0)\mid
B_s^*(q+p)\rangle\langle B^*_s| B_{s1}\eta\rangle  \langle
B_{s1}(q)|{J^A_{\nu}}^\dagger(0)| 0\rangle}
  {\left[M_{B_s^*}^2-(q+p)^2\right]\left[M_{B_{s1}}^2-q^2\right]}   \nonumber \\
  &&+\frac{\langle0| J^V_{\mu}(0)\mid
B_s^*(q+p)\rangle\langle B^*_s| B_{s}\eta\rangle  \langle
B_{s}(q)|{J^A_{\nu}}^\dagger(0)| 0\rangle}
  {\left[M_{B_s^*}^2-(q+p)^2\right]\left[M_{B_{s}}^2-q^2\right]}    \nonumber \\
 && +\frac{\langle0| J^V_{\mu}(0)\mid
B_{s0}(q+p)\rangle\langle B_{s0}| B_{s}\eta\rangle  \langle
B_{s}(q)|{J^A_{\nu}}^\dagger(0)| 0\rangle}
  {\left[M_{B_{s0}}^2-(q+p)^2\right]\left[M_{B_{s}}^2-q^2\right]}  + \cdots \, ,\nonumber \\
 &=&-\frac{i g_{B_{s1}B^*_s \eta} f_{B^*_s} f_{B_{s1}}
M_{B^*_s}M_{B_{s1}}}
  {\left[M_{B_s^*}^2-(q+p)^2\right]\left[M_{B_{s1}}^2-q^2\right]}\left[
  -g_{\mu\lambda}+\frac{(p+q)_\mu(p+q)_\lambda}{M_{B_s^*}^2} \right]
   \nonumber \\
  && \left[-g_{\lambda\nu}+\frac{q_\lambda q_\nu}{M_{B_{s1}}^2} \right]+iC_1   \left[
  -g_{\mu\lambda}+\frac{(p+q)_\mu(p+q)_\lambda}{M_{B_s^*}^2} \right]p^\lambda q_\nu  \nonumber\\
 && +iC_2 (p+q)_\mu q_\nu   + \cdots \, , \nonumber \\
 &=&-\frac{i g_{B_{s1}B^*_s \eta} f_{B^*_s} f_{B_{s1}}
M_{B^*_s}M_{B_{s1}}}
  {\left[M_{B_s^*}^2-(q+p)^2\right]\left[M_{B_{s1}}^2-q^2\right]}g_{\mu\nu}+\cdots
  \, ,
 \end{eqnarray}

 \begin{eqnarray}
 \Pi_{\mu}&=&\frac{\langle0| J^5_{\mu}(0)\mid B_s(q+p)\rangle\langle
B_s| B_{s0}\eta\rangle  \langle B_{s0}(q)|J^{S\dagger}(0)| 0\rangle}
  {\left[M_{B_s}^2-(q+p)^2\right]\left[M_{B_{s0}}^2-q^2\right]}  \nonumber \\
  &&+\frac{\langle0| J^5_{\mu}(0)\mid B_{s1}(q+p)\rangle\langle
B_{s1}| B_{s0}\eta\rangle  \langle B_{s0}(q)|J^{S\dagger}(0)|
0\rangle}
  {\left[M_{B_{s1}}^2-(q+p)^2\right]\left[M_{B_{s0}}^2-q^2\right]}  + \cdots \, ,\nonumber \\
  &=&\frac{i g_{B_{s0}B_s\eta} f_{B_s} f_{B_{s0}}  M_{B_{s0}}}
  {\left[M_{B_s}^2-(q+p)^2\right]\left[M_{B_{s0}}^2-q^2\right]}(p+q)_{\mu}
  \nonumber \\
  &&+iC_3 \left[ -g_{\mu\lambda}+\frac{(p+q)_\mu(p+q)_\lambda}{M_{B_{s1}}^2} \right] p_\lambda  + \cdots \, ,\nonumber \\
  &=&\frac{i g_{B_{s0}B_s\eta} f_{B_s} f_{B_{s0}}  M_{B_{s0}}}
  {\left[M_{B_s}^2-(q+p)^2\right]\left[M_{B_{s0}}^2-q^2\right]}q_{\mu} +
  iC_3 \frac{M_{B^*_{s}}^2+m_\eta^2-M_{B_{s1}}^2}{2M_{B_{s1}}^2} q_\mu  +
  \cdots \, ,
  \end{eqnarray}
where the following definitions for the weak decay constants have
been used,
\begin{eqnarray}
\langle0 | J^V_{\mu}(0)|B_s^*(p)\rangle&=&f_{B^*_s}M_{B^*_s}\epsilon_\mu\,, \nonumber\\
\langle0|J^A_{\mu}(0)|B_{s1}(p)\rangle&=&f_{B_{s1}}M_{B_{s1}}\eta_\mu\,,\nonumber \\
\langle0 | J^5_{\mu}(0)|B_s(p)\rangle&=&if_{B_s}p_\mu\,, \nonumber\\
\langle0 | J^S(0)|B_{s0}(p)\rangle&=&f_{B_{s0}}M_{B_{s0}} \,,\nonumber\\
 \langle0 |J^V_{\mu}(0)|B_{s0}(p)\rangle&=&f_{B_{s0}}p_\mu \, .
\end{eqnarray}
We introduce the notations $C_i$ for simplicity, the explicit
expressions are neglected as the contributions can be deleted with
suitable tensor structures. The term proportional to the $C_3$ is
greatly suppressed by the small numerical factor
$\frac{M_{B^*_{s}}^2+m_\eta^2-M_{B_{s1}}^2}{M_{B_{s1}}^2}$, and the
contributions  from the axial-vector meson can be neglected safely
in Eq.(8). We choose the tensor structure $g_{\mu\nu}$ to avoid the
contaminations from the scalar meson $B_{s0}$ and the pseudoscalar
meson $B_s$ in the sum rule for the strong coupling constant
$g_{B_{s1} B^*_s \eta}$.  In deriving the sum rule for the strong
coupling constant $g_{B_{s0} B_s \eta}$, we choose the axial-vector
current $J^5_\mu(x)$ to interpolate the pseudoscalar meson $B_s$,
although there are contaminations from the axial-vector meson
$B_{s1}$,  the contaminations are tiny  and can be neglected safely
if we choose the tensor structure $q_\mu$. If we choose the
pseudoscalar current $J_5(x)=\bar{s}(x) i \gamma_5 c(x)$ to
interpolate the pseudoscalar meson $B_s$, the axial-vector mesons
have no contaminations,  I fail to take notice of this fact at
beginning of the work.

We perform the  operator product expansion for the correlation
functions $\Pi_{\mu \nu}(p,q)$ and $\Pi_{\mu }(p,q)$ in perturbative
QCD theory,  and  obtain the analytical expressions at the level of
quark-gluon degrees of freedom. In calculation, the  two-particle
and three-particle $\eta$ meson light-cone distribution amplitudes
have been used \cite{PSLC1,PSLC2}, the explicit expressions are
given in the appendix. The parameters in the light-cone distribution
amplitudes are scale dependent and are calculated  with the QCD sum
rules \cite{PSLC1,PSLC2}. In this article, the energy scale $\mu$ is
chosen to be $\mu=1\rm{GeV}$,  one can choose another typical energy
scale $\mu=\sqrt{M_B^2-m_b^2}\approx 2.4\rm{GeV}$. The light-cone
distribution amplitudes are calculated at the energy scale
$\mu=1\rm{GeV}$ with the QCD sum rules,  evolution of  the
coefficients to larger energy scales with the (complex)
re-normalization group equation which concerns approximations in one
or other ways, additional uncertainties are introduced. The physical
quantities would not depend on the special energy scale we choose,
we expect that scale dependence of the input parameters is canceled
out approximately with each other, the values of the strong coupling
constants which are calculated  at the energy scale $\mu=1\rm{GeV}$
can make robust predictions. Furthermore, in the heavy quark limit,
the bound energy of the strange-bottomed $(0^+,1^+)$ mesons is about
$\Lambda=\frac{3M_{B_{s1}}+M_{B_{s0}}}{4}-m_b\approx 1\rm{GeV}$,
which can serve as a typical energy scale and validate our choice.

After straightforward calculations, we obtain the final expressions
of the double Borel transformed correlation functions $\Pi_A $ and
$\Pi_S$ at the level of quark-gluon degrees of freedom. The masses
of  the strange-bottomed mesons are $M_{B_{s1}}=5.72\rm{GeV}$,
$M_{B_{s0}}=5.70\rm{GeV}$, $M_{B^*_s}=5.412\rm{GeV}$ and
$M_{B_s}=5.366\rm{GeV}$,
\begin{eqnarray}
 \frac{M_{B_{s1}}^2}{M_{B_{s1}}^2+M_{B_s^*}^2}\approx
\frac{M_{B_{s0}}^2}{M_{B_{s0}}^2+M_{B_s}^2}\approx0.53 \, ,
\end{eqnarray}
 there exists an overlapping working window for the two Borel
parameters $M_1^2$ and $M_2^2$, it's convenient to take the value
$M_1^2=M_2^2$. We introduce the threshold parameter $s_0$ (denotes
$s_S^0$ and $s_A^0$) and make the simple replacement,
\begin{eqnarray}
e^{-\frac{m_b^2+u_0(1-u_0)m_\eta^2}{M^2}} \rightarrow
e^{-\frac{m_b^2+u_0(1-u_0)m_\eta^2}{M^2} }-e^{-\frac{s_0}{M^2}}
\nonumber
\end{eqnarray}
 to subtract the contributions from the high resonances  and
  continuum states \cite{Belyaev94}, finally we obtain the sum rules  for the strong coupling
  constants $g_{B_{s0}B_s \eta}$ and
$g_{B_{s1}B^*_s \eta}$ respectively\footnote{For example, we use the
notation  $(A_\parallel+A_\perp)(1-\alpha-\beta,\alpha,\beta)$ to
represent
$A_\parallel(1-\alpha-\beta,\alpha,\beta)+A_\perp(1-\alpha-\beta,\alpha,\beta)$.
Other expressions can be understood  in the same way.  },

\begin{eqnarray}
g_{B_{s0}B_s \eta}&=&
\frac{1}{f_{B_s}f_{B_{s0}}M_{B_{s0}}}\exp\left(
\frac{M^2_{B_{s0}}}{M_1^2} +\frac{M^2_{B_s}}{M_2^2}
\right)\left\{\left[\exp\left(- \frac{\Xi}{M^2}\right)-\exp\left(-
\frac{s_S^0}{M^2}\right)\right]  \right.\nonumber\\
&& \frac{f'_\eta
m_\eta^2M^2}{m_s}\left[\varphi_p(u_0)-\frac{d\varphi_\sigma(u_0)}{6du_0}\right]
+\exp\left(-\frac{\Xi}{M^2}\right)\left[ -m_bf'_\eta m_\eta^2
\int_0^{u_0}
dt B(t)  \right.\nonumber\\
  &&+f'_{3\eta}m_\eta^2 \int_0^{u_0}d\alpha_s
  \int_{u_0-\alpha_s}^{1-\alpha_s}d\alpha_g
\varphi_{3\eta}(1-\alpha_s-\alpha_g,\alpha_g,\alpha_s)\frac{2(\alpha_s+\alpha_g-u_0)-3\alpha_g
}{\alpha_g^2}
\nonumber\\
&&-\frac{2m_bf'_\eta m_\eta^4}{M^2}  \int_{1-u_0}^1 d\alpha_g
\frac{1-u_0}{\alpha_g^2}\int_0^{\alpha_g}
d\beta\int_0^{1-\beta}d\alpha \Phi(1-\alpha-\beta,\beta,\alpha)
\nonumber \\
&& +\frac{2m_bf'_\eta m_\eta^4}{M^2}\left(\int_0^{1-u_0} d\alpha_g
\int^{u_0}_{u_0-\alpha_g} d\alpha_s \int_0^{\alpha_s} d\alpha
+\int^1_{1-u_0} d\alpha_g \int^{1-\alpha_g}_{u_0-\alpha_g} d\alpha_s
\int_0^{\alpha_s} d\alpha\right) \nonumber\\
&&\left.\left.\frac{\Phi(1-\alpha-\alpha_g,\alpha_g,\alpha)}{\alpha_g}
\right]\right\} \, ,
\end{eqnarray}

\begin{eqnarray}
 g_{B_{s1}B^*_s \eta}&=&
\frac{1}{f_{B_s^*}f_{B_{s1}}M_{B^*_s}M_{B_{s1}}} \exp\left(
\frac{M^2_{B_{s1}}}{M_1^2} +\frac{M^2_{B_s^*}}{M_2^2} \right)
\left\{\left[\exp\left(- \frac{\Xi}{M^2}\right)-\exp\left(-
\frac{s^0_A}{M^2}\right)\right] \right.\nonumber\\
 &&f'_\eta\left[\frac{ m_b m_\eta^2M^2}{m_s}
\varphi_p(u_0)
+\frac{m_\eta^2(M^2+m_b^2)}{8}  \frac{d}{du_0}A(u_0) -\frac{M^4}{2} \frac{d}{du_0}\phi_\eta(u_0)\right] \nonumber\\
&&-\exp\left(-\frac{\Xi}{M^2}\right) \left[f'_\eta m_b^2 m_\eta^2
\int_0^{u_0} dt B(t) \right.\nonumber\\
&&+ m_\eta^2 \int_0^{u_0} d\alpha_s
\int_{u_0-\alpha_s}^{1-\alpha_s} d\alpha_g \frac{(u_0f'_\eta m_\eta^2\Phi+f'_{3\eta}m_b \varphi_{3\eta})(1-\alpha_s-\alpha_g,\alpha_s,\alpha_g)}{\alpha_g} \nonumber \\
&&+f'_\eta m_\eta^2M^2\frac{d}{du_0}\int_0^{u_0} d\alpha_s
\int_{u_0-\alpha_s}^{1-\alpha_s} d\alpha_g
\frac{(A_\parallel-V_\parallel)(1-\alpha_s-\alpha_g,\alpha_s,\alpha_g)}{2\alpha_g}
 \nonumber \\
&&-f'_\eta m_\eta^2M^2\frac{d}{du_0}\int_0^{u_0} d\alpha_s
\int_{u_0-\alpha_s}^{1-\alpha_s} d\alpha_g
A_\parallel(1-\alpha_s-\alpha_g,\alpha_s,\alpha_g)\frac{\alpha_s+\alpha_g-u_0}{\alpha_g^2}
 \nonumber \\
 &&+f'_\eta m_\eta^4  \left(\int_0^{1-u_0} d\alpha_g \int^{u_0}_{u_0-\alpha_g}
d\alpha_s \int_0^{\alpha_s} d\alpha +\int^1_{1-u_0} d\alpha_g
\int^{1-\alpha_g}_{u_0-\alpha_g}
d\alpha_s \int_0^{\alpha_s} d\alpha\right)  \nonumber \\
&& \left[\frac{1}{\alpha_g}\left(3-\frac{2m_b^2}{M^2}\right)\Phi+\frac{4m_b^2}{M^2}\frac{\alpha_s+\alpha_g-u_0}{\alpha_g^2}(A_\perp+A_\parallel)\right](1-\alpha-\alpha_g,\alpha,\alpha_g)\nonumber \\
 &&-f'_\eta m_\eta^4  u_0\frac{d}{du_0}\left(\int_0^{1-u_0} d\alpha_g \int^{u_0}_{u_0-\alpha_g}
d\alpha_s \int_0^{\alpha_s} d\alpha +\int^1_{1-u_0} d\alpha_g
\int^{1-\alpha_g}_{u_0-\alpha_g}
d\alpha_s \int_0^{\alpha_s} d\alpha\right)  \nonumber \\
&& \frac{\Phi(1-\alpha-\alpha_g,\alpha,\alpha_g)}{\alpha_g} \nonumber \\
&&-f'_\eta m_\eta^4   \int_{1-u_0}^1 d\alpha_g \int_0^{\alpha_g}
d\beta \int_0^{1-\beta} d\alpha
\left[\Phi(1-\alpha-\beta,\alpha,\beta) \frac{1-u_0}{\alpha_g^2}
 \left(4-\frac{2m_b^2}{M^2}\right) \right.\nonumber \\
&&\left. +\frac{4m_b^2}{M^2}\frac{(1-u_0)^2}{\alpha_g^3}(A_\parallel+A_\perp)(1-\alpha-\beta,\alpha,\beta)\right]\nonumber \\
&&\left.\left.+f'_\eta m_\eta^4  \frac{d}{du_0} \int_{1-u_0}^1
d\alpha_g \int_0^{\alpha_g} d\beta \int_0^{1-\beta} d\alpha
\Phi(1-\alpha-\beta,\alpha,\beta) \frac{u_0(1-u_0)}{\alpha_g^2}
\right]\right\} ,
\end{eqnarray}
where
\begin{eqnarray}
\Phi(\alpha_i)&=&A_\parallel(\alpha_i)+A_\perp(\alpha_i)-V_\parallel(\alpha_i)-V_\perp(\alpha_i) \, ,\nonumber \\
\Xi&=&m_b^2+u_0(1-u_0)m_\eta^2 \, ,\nonumber \\
u_0&=&\frac{M_1^2}{M_1^2+M_2^2}\, , \nonumber \\
M^2&=&\frac{M_1^2M_2^2}{M_1^2+M_2^2} \, .
\end{eqnarray}

\section{Numerical result and discussion}
The input parameters are taken as $m_s=(140\pm 10 )\rm{MeV}$,
 $m_b=(4.7\pm 0.1)\rm{GeV}$,
$\lambda_3=0.0$, $a_1=0.0$, $f_{3\eta}=(0.40\pm0.12)\times
10^{-2}\rm{GeV}^2$, $\omega_3=-3.0\pm0.9$, $\eta_4=0.5 \pm0.2 $,
$\omega_4=0.2\pm0.1$, $a_2=0.20\pm 0.06$ \cite{PSLC1,PSLC2},
$f_\eta=0.145\rm{GeV}$, $m_\eta=0.548\rm{GeV}$,
$f'_\eta=-\frac{2}{\sqrt{6}}f_\eta$,
$f'_{3\eta}=-\frac{2}{\sqrt{6}}f_{3\eta}$, $M_{B_s}=5.366\rm{GeV}$,
$M_{B^*_s}=5.412\rm{GeV}$ \cite{PDG},
$M_{B_{s0}}=(5.70\pm0.11)\rm{GeV} $,
$M_{B_{s1}}=(5.72\pm0.09)\rm{GeV}$, $f_{B_{s0}}= f_{B_{s1}}=(0.24\pm
0.02)\rm{GeV}$ \cite{Wang0712},
$f_{B^*_s}=f_{B_s}=(0.19\pm0.02)\rm{GeV}$
\cite{LCSRreview,Wang04BS,Verde-Velasco07},
  $s^0_S=(37 \pm 1)\rm{GeV}^2$ and $s^0_A=(38 \pm 1)\rm{GeV}^2$ \cite{Wang0712}.
  The Borel parameters are chosen as $M^2=(5-7)\rm{GeV}^2$, in this region, the
values of the strong coupling constants  $g_{B_{s1} B_s^* \eta}$ and
$g_{B_{s0} B_s \eta}$ are rather stable, which are shown in Fig.1.
\begin{figure} \centering
  \includegraphics[totalheight=6cm,width=7cm]{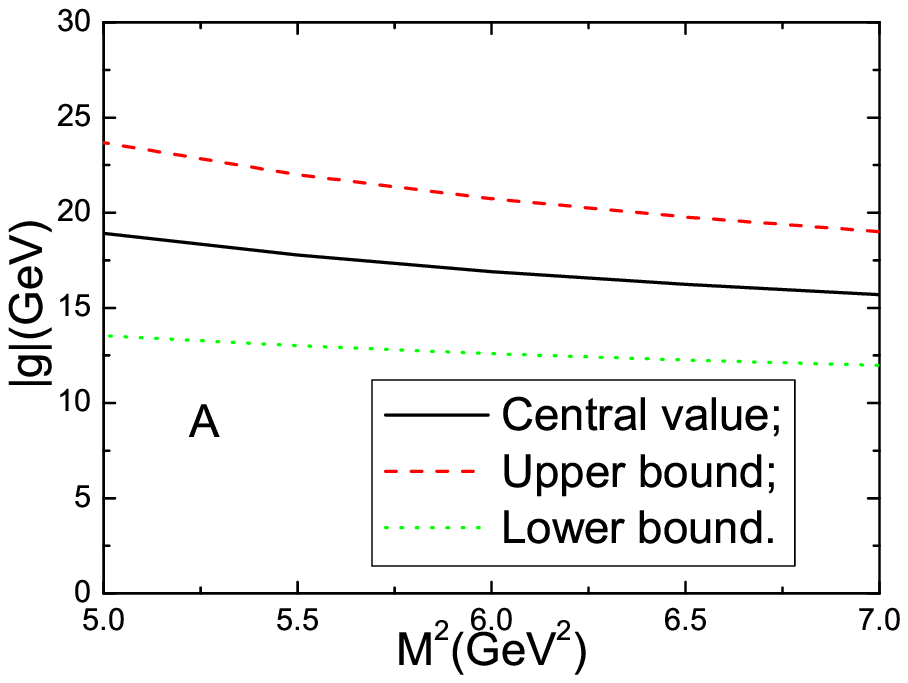}
 \includegraphics[totalheight=6cm,width=7cm]{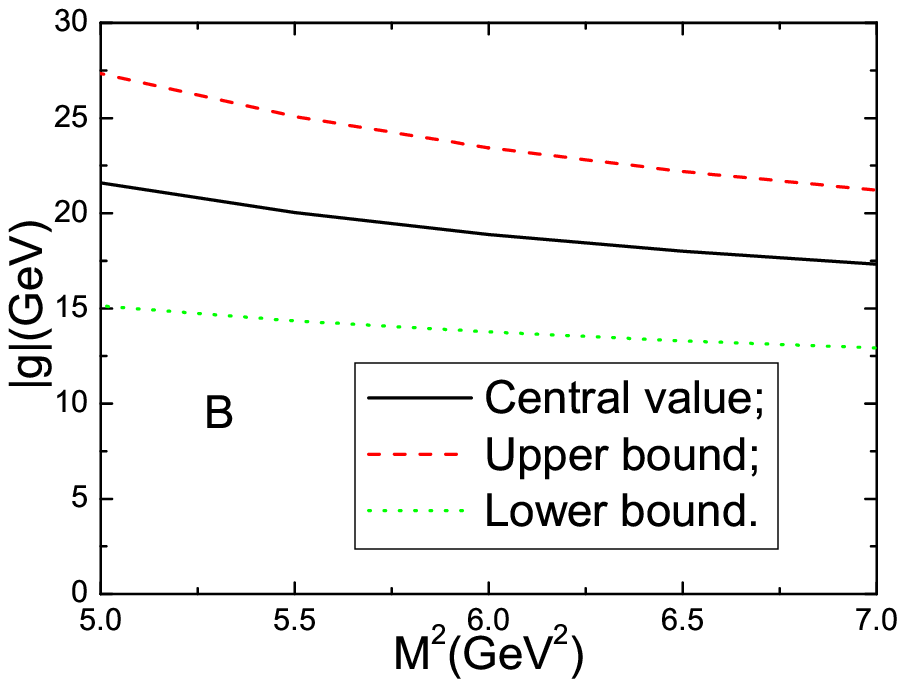}
     \caption{The strong coupling constants  $g_{B_{s1}B^*_s \eta}$(A) and  $g_{B_{s0}B_s \eta}$(B) with the parameter $M^2$. }
\end{figure}

In the limit of large Borel parameter $M^2$, the strong coupling
constants $g_{B_{s1} B^*_s \eta}$ and $g_{B_{s0} B_s \eta}$ take up
the following behaviors respectively,
\begin{eqnarray}
g_{B_{s0}B_s \eta}&\propto& \frac{
M^2\varphi_p(u_0)}{f_{B_s}f_{B_{s0}}}
\, , \nonumber \\
g_{B_{s1}B^*_s \eta}&\propto& \frac{m_b
M^2\varphi_p(u_0)}{f_{B^*_s}f_{B_{s1}}} \, .
\end{eqnarray}
It is not unexpected, the contributions  from the two-particle
twist-3 light-cone distribution amplitude $\varphi_p(u)$ are greatly
enhanced by the large Borel parameter $M^2$,  (large) uncertainties
of the relevant parameters presented in above equations have
significant impact on the numerical results. The contribution from
the two-particle twist-3 light-cone distribution amplitude
$\varphi_\sigma(u_0)$ is zero due to symmetry property.

Taking into account all the uncertainties of the input parameters,
finally we obtain the numerical values of the strong coupling
constants, which are shown in Fig.1,
\begin{eqnarray}
  |g_{B_{s1}B_s^* \eta}| &=&(17.8\pm5.8) \rm{GeV} \, , \nonumber\\
  |g_{B_{s0}B_s \eta}| &=&(20.1\pm7.2) \rm{GeV} \, ,
 \end{eqnarray}
the uncertainties are large, about $30\%$.  Taking into account the
 small $\eta-\pi^0$ transition matrix   according to
   Dashen's
 theorem \cite{Dashen}$,
 t_{\eta\pi} = \langle \pi^0 |\mathcal {H}
 |\eta\rangle=-0.003\rm{GeV}^2$, we can obtain
  the narrow decay widths.
\begin{eqnarray}
  \Gamma_{B_{s1}B_s^* \pi} &=& \frac{p_1}{24\pi M_{B_{s1}}^2} \sum_\lambda \sum_{\lambda'}\mid\frac{g_{B_{s1}B_s^* \eta}\eta^*(\lambda)\cdot \epsilon(\lambda') t_{\eta\pi} }{m_\pi^2-m_\eta^2}\mid^2=(5.3-20.7) \rm{KeV} \, , \nonumber\\
  \Gamma_{B_{s0}B_s \pi} &=&\frac{p_2}{8\pi M_{B_{s0}}^2}\mid\frac{g_{B_{s0}B_{s}\eta}t_{\eta\pi} }{m_\pi^2-m_\eta^2}\mid^2= (6.8-30.7) \rm{KeV} \, ,\\
  p_1&=&\frac{\sqrt{\left[M_{B_{s1}}^2-(M_{B^*_s}+m_\pi)^2\right]\left[M_{B_{s1}}^2-(M_{B^*_s}-m_\pi)^2\right]}}{2M_{B_{s1}}}
  \, ,\nonumber \\
  p_2&=&\frac{\sqrt{\left[M_{B_{s0}}^2-(M_{B_s}+m_\pi)^2\right]\left[M_{B_{s0}}^2-(M_{B_s}-m_\pi)^2\right]}}{2M_{B_{s0}}}
  \, , \nonumber
 \end{eqnarray}
which are consistent with the ones obtained from the analysis of the
unitarized two-meson scattering amplitudes with the heavy-light
chiral lagrangian, $\Gamma_{B_{s1}B_s^* \pi} = 10.36\rm{KeV} $ and $
\Gamma_{B_{s0}B_s \pi} =7.92 \rm{KeV}$ \cite{GuoD1,GuoD2}.  We can
search the strange-bottomed $(0^+,1^+)$ mesons $B_{s0}$ and $B_{s1}$
in the invariant $B_s \pi^0$ and $B^*_s \pi^0$ mass distributions
respectively, just like the BaBar and CLEO Collaborations observed
the strange-charmed $(0^+,1^+)$ mesons $D_{s0}$ and $D_{s1}$ in the
invariant $D_s \pi^0$ and $D^*_s \pi^0$ mass distributions
respectively \cite{Babar03,CLEO03}.

\section{Conclusion}

In this article, we calculate the strong coupling constants
$g_{B_{s0} B_s \eta}$ and $g_{B_{s1} B^*_s \eta}$ with the
light-cone QCD sum rules. Then we take into account the small
$\eta-\pi^0$ transition matrix   according to   Dashen's
 theorem, and obtain the small decay widths. We can search the strange-bottomed
$(0^+,1^+)$ mesons $B_{s0}$ and $B_{s1}$ in the  invariant $B_s
\pi^0$ and $B^*_s \pi^0$ mass distributions respectively.

\section*{Acknowledgments}
This  work is supported by National Natural Science Foundation,
Grant Number 10405009, 10775051, and Program for New Century
Excellent Talents in University, Grant Number NCET-07-0282.

 \section*{Appendix}
 The light-cone distribution amplitudes of the $\eta$ meson are defined
 by
\begin{eqnarray}
\langle0| {\bar s} (0) \gamma_\mu \gamma_5 s(x) |\eta(p)\rangle& =&
i f'_\eta p_\mu \int_0^1 du  e^{-i u p\cdot x}
\left\{\phi_\eta(u)+\frac{m_\eta^2x^2}{16}
A(u)\right\}\nonumber\\
&&+f'_\eta m_\eta^2\frac{ix_\mu}{2p\cdot x}
\int_0^1 du  e^{-i u p \cdot x} B(u) \, , \nonumber\\
\langle0| {\bar s} (0) i \gamma_5 s(x) |\eta(p)\rangle &=&
\frac{f'_\eta m_\eta^2}{
m_s}\int_0^1 du  e^{-i u p \cdot x} \varphi_p(u)  \, ,  \nonumber\\
\langle0| {\bar s} (0) \sigma_{\mu \nu} \gamma_5 s(x)
|\eta(p)\rangle &=&i(p_\mu x_\nu-p_\nu x_\mu)  \frac{f'_\eta
m_\eta^2}{6 m_s} \int_0^1 du
e^{-i u p \cdot x} \varphi_\sigma(u) \, ,  \nonumber\\
\langle0| {\bar s} (0) \sigma_{\alpha \beta} \gamma_5 g_s G_{\mu
\nu}(v x)s(x) |\eta(p)\rangle&=& f'_{3 \eta}\left\{(p_\mu p_\alpha
g^\bot_{\nu
\beta}-p_\nu p_\alpha g^\bot_{\mu \beta}) -(p_\mu p_\beta g^\bot_{\nu \alpha}\right.\nonumber\\
&&\left.-p_\nu p_\beta g^\bot_{\mu \alpha})\right\} \int {\cal
D}\alpha_i \varphi_{3 \eta} (\alpha_i)
e^{-ip \cdot x(\alpha_s+v \alpha_g)} \, ,\nonumber\\
\langle0| {\bar s} (0) \gamma_{\mu} \gamma_5 g_s G_{\alpha
\beta}(vx)s(x) |\eta(p)\rangle&=&  p_\mu  \frac{p_\alpha
x_\beta-p_\beta x_\alpha}{p
\cdot x}f'_\eta m_\eta^2\nonumber\\
&&\int{\cal D}\alpha_i A_{\parallel}(\alpha_i) e^{-ip\cdot
x(\alpha_s +v \alpha_g)}\nonumber \\
&&+ f'_\eta m_\eta^2 (p_\beta g_{\alpha\mu}-p_\alpha
g_{\beta\mu})\nonumber\\
&&\int{\cal D}\alpha_i A_{\perp}(\alpha_i)
e^{-ip\cdot x(\alpha_s +v \alpha_g)} \, ,  \nonumber\\
\langle0| {\bar s} (0) \gamma_{\mu}  g_s \tilde G_{\alpha
\beta}(vx)s(x) |\eta(p)\rangle&=& p_\mu  \frac{p_\alpha
x_\beta-p_\beta x_\alpha}{p \cdot
x}f'_\eta m_\eta^2\nonumber\\
&&\int{\cal D}\alpha_i V_{\parallel}(\alpha_i) e^{-ip\cdot
x(\alpha_s +v \alpha_g)}\nonumber \\
&&+ f'_\eta m_\eta^2 (p_\beta g_{\alpha\mu}-p_\alpha
g_{\beta\mu})\nonumber\\
&&\int{\cal D}\alpha_i V_{\perp}(\alpha_i) e^{-ip\cdot x(\alpha_s +v
\alpha_g)} \, ,
\end{eqnarray}
where the operator $\tilde G_{\alpha \beta}$  is the dual of the
$G_{\alpha \beta}$, $\tilde G_{\alpha \beta}= {1\over 2}
\epsilon_{\alpha \beta  \mu\nu} G^{\mu\nu} $ and ${\cal{D}}\alpha_i$
is defined as ${\cal{D}} \alpha_i =d \alpha_{\bar{s}} d \alpha_g d
\alpha_s \delta(1-\alpha_{\bar{s}} -\alpha_g -\alpha_s)$. The
light-cone distribution amplitudes are parameterized as
\begin{eqnarray}
\phi_\eta(u)&=&6u(1-u)
\left\{1+a_1C^{\frac{3}{2}}_1(2u-1)+a_2C^{\frac{3}{2}}_2(2u-1)\right\}\, , \nonumber\\
\varphi_p(u)&=&1+\left\{30\eta_3-\frac{5}{2}\rho^2\right\}C_2^{\frac{1}{2}}(2u-1)\nonumber \\
&&+\left\{-3\eta_3\omega_3-\frac{27}{20}\rho^2-\frac{81}{10}\rho^2 a_2\right\}C_4^{\frac{1}{2}}(2u-1)\, ,  \nonumber \\
\varphi_\sigma(u)&=&6u(1-u)\left\{1
+\left[5\eta_3-\frac{1}{2}\eta_3\omega_3-\frac{7}{20}\rho^2-\frac{3}{5}\rho^2 a_2\right]C_2^{\frac{3}{2}}(2u-1)\right\}\, , \nonumber \\
\varphi_{3\eta}(\alpha_i) &=& 360 \alpha_{\bar{s}} \alpha_s
\alpha_g^2 \left \{1 +\lambda_3(\alpha_{\bar{s}}-\alpha_s)+ \omega_3
\frac{1}{2} ( 7 \alpha_g
- 3) \right\} \, , \nonumber\\
V_{\parallel}(\alpha_i) &=& 120\alpha_{\bar{s}} \alpha_s \alpha_g
\left( v_{00}+v_{10}(3\alpha_g-1)\right)\, ,
\nonumber \\
A_{\parallel}(\alpha_i) &=& 120 \alpha_{\bar{s}} \alpha_s \alpha_g
a_{10} (\alpha_s-\alpha_{\bar{s}})\, ,
\nonumber\\
V_{\perp}(\alpha_i) &=& -30\alpha_g^2
\left\{h_{00}(1-\alpha_g)+h_{01}\left[\alpha_g(1-\alpha_g)-6\alpha_{\bar{s}}
\alpha_s\right] \right.  \nonumber\\
&&\left. +h_{10}\left[
\alpha_g(1-\alpha_g)-\frac{3}{2}\left(\alpha_{\bar{s}}^2+\alpha_s^2\right)\right]\right\}\,
, \nonumber\\
A_{\perp}(\alpha_i) &=&  30 \alpha_g^2 (\alpha_{\bar{s}}-\alpha_s)
\left\{h_{00}
+h_{01}\alpha_g+\frac{1}{2}h_{10}(5\alpha_g-3)  \right\}, \nonumber\\
A(u)&=&6u(1-u)\left\{
\frac{16}{15}+\frac{24}{35}a_2+20\eta_3+\frac{20}{9}\eta_4 \right.
\nonumber \\
&&+\left[
-\frac{1}{15}+\frac{1}{16}-\frac{7}{27}\eta_3\omega_3-\frac{10}{27}\eta_4\right]C^{\frac{3}{2}}_2(2u-1)
\nonumber\\
&&\left.+\left[
-\frac{11}{210}a_2-\frac{4}{135}\eta_3\omega_3\right]C^{\frac{3}{2}}_4(2u-1)\right\}+\left\{
 -\frac{18}{5}a_2+21\eta_4\omega_4\right\} \nonumber\\
 && \left\{2u^3(10-15u+6u^2) \log u+2\bar{u}^3(10-15\bar{u}+6\bar{u}^2) \log \bar{u}
 \right. \nonumber\\
 &&\left. +u\bar{u}(2+13u\bar{u})\right\} \, ,\nonumber\\
 g_\eta(u)&=&1+g_2C^{\frac{1}{2}}_2(2u-1)+g_4C^{\frac{1}{2}}_4(2u-1)\, ,\nonumber\\
 B(u)&=&g_\eta(u)-\phi_\eta(u)\, ,
\end{eqnarray}
where
\begin{eqnarray}
h_{00}&=&v_{00}=-\frac{\eta_4}{3} \, ,\nonumber\\
a_{10}&=&\frac{21}{8}\eta_4 \omega_4-\frac{9}{20}a_2 \, ,\nonumber\\
v_{10}&=&\frac{21}{8}\eta_4 \omega_4 \, ,\nonumber\\
h_{01}&=&\frac{7}{4}\eta_4\omega_4-\frac{3}{20}a_2 \, ,\nonumber\\
h_{10}&=&\frac{7}{2}\eta_4\omega_4+\frac{3}{20}a_2 \, ,\nonumber\\
g_2&=&1+\frac{18}{7}a_2+60\eta_3+\frac{20}{3}\eta_4 \, ,\nonumber\\
g_4&=&-\frac{9}{28}a_2-6\eta_3\omega_3 \, ,
\end{eqnarray}
 here  $ C_2^{\frac{1}{2}}(\xi)$, $ C_4^{\frac{1}{2}}(\xi)$
 and $ C_2^{\frac{3}{2}}(\xi)$ are Gegenbauer polynomials,
  $\eta_3=\frac{f_{3\eta}}{f_\eta}\frac{m_s}{m_\eta^2}$ and  $\rho^2={m_s^2\over m_\eta^2}$
 \cite{PSLC1,PSLC2}.

\end{document}